\documentstyle[12pt]{article}  
\def\beq{\begin{equation}}
\def\eeq{\end{equation}}

\def\bea{\begin{eqnarray}}
\def\eea{\end{eqnarray}}
\def\eq#1{eq.~(\ref{#1})}
\def\mp{M_{\rm Pl}}

\begin{document}

\baselineskip=18pt
\begin{center}
{\Large \bf NATURALLY SPEAKING: \\
The Naturalness Criterion and Physics at the LHC}  

Gian Francesco GIUDICE

CERN, Theoretical Physics Division\\
Geneva, Switzerland

\end{center}

\section{Naturalness in Scientific Thought}
\begin{flushright}
{\it Everything is natural: if it weren't, it wouldn't be.}\\
Mary Catherine Bateson~\cite{bateson}
\end{flushright}

Almost every branch of science has its own version of the ``naturalness criterion". In environmental sciences, it refers to the degree to which an area is pristine, free from human influence, and characterized by native species~\cite{anders}. In mathematics, its meaning is associated with the intuitiveness of certain fundamental concepts, viewed as an intrinsic part of our thinking~\cite{natmat}. One can find the use of naturalness criterions in computer science (as a measure of adaptability), in agriculture (as an acceptable level of product manipulation), in linguistics (as translation quality assessment of sentences that do not reflect the natural and idiomatic forms of the receptor language). But certainly nowhere else but in particle physics has the mutable concept of naturalness taken a form which has become so influential in the development of the field.

The role of naturalness in the sense of ``{\ae}sthetic beauty" is a powerful guiding principle for physicists as they try to construct new theories. This may appear surprising since the final product is often a mathematically sophisticated theory based on deep fundamental principles, and one could believe that subjective {\ae}sthetic arguments have no place in it. Nevertheless, this is not true and often theoretical physicists formulate their theories inspired by criteria of simplicity and beauty, {\it i.e.} by what Nelson~\cite{nelson} defines as ``structural naturalness". When Einstein was asked what he would have done, had Eddington's observation of the 1919 solar eclipse disproved, rather than confirmed, his theory, he simply replied: {\it ``Then I would have felt sorry for the dear Lord"}~\cite{einquo}. Clearly he was confident that the structural naturalness of general relativity was no frippery.

Structural naturalness is a powerful inspirational principle but, of course, it cannot be used to validate a theory. Moreover, since it is subjected to philosophical influences and to the limited scientific knowledge of the time, sometimes it can even be misleading. From a modern point of view, the solar system is more {\it naturally} explained by a heliocentric theory, in which planetary motions are described by simple elliptic orbits, rather than by a geocentric theory, which requires the introduction of different epicycles for each planet. But to predecessors and contemporaries of Copernicus a geocentric theory probably appeared more {\it natural}. Tycho Brahe discarded a heliocentric description of the solar system with the harsh, but rather unconvincing, argument that the Earth is a {\it ``hulking, lazy body, unfit for motion"}~\cite{brahe}. Certainly Aristotelian and biblical influences had their part in forming this belief, but a big role was played by the incorrect scientific notion that we would be able to feel the Earth moving under our feet. 

Aristarchus of Samos was the first to postulate that the Sun was at the center of the universe, but
the ancient Greeks ruled out the heliocentric model based on the following ``naturalness" argument. 
Assuming proportionality between the period and the radius of planetary orbits, they obtained that Saturn is 29 times as far from the Sun than the Earth, since the period of Saturn was known to be 29 years. Using trigonometry and some astronomical observations, Aristarchus obtained the Sun-Earth distance expressed in terms of the Earth radius $R_\oplus$ previously deduced by Erathostenes with his famous measurement of the inclination of the solar rays in Alexandria when the Sun was at zenith in Syene. This placed Saturn at a distance of 20,000 $R_\oplus$ from the Earth\footnote{The modern value of the minimum distance between Saturn and Earth is $1.9\times 10^5~R_\oplus$.}~\cite{cushing}. Since Saturn was the outermost known planet, it was {\it natural} to assume that the universe was about the same size. But if the Earth orbits around the Sun, we should observe a parallax effect for stars on a celestial sphere of radius 20,000 $R_\oplus$. No stellar parallax could be observed with naked eye (for Alpha Centauri, the closest star, the parallax angle is actually only about one second of arc), and the heliocentric model was rejected. Copernicus dispensed with the parallax objection by refuting the {\it natural} assumption about stellar distances and required that stars be at least 1,500,000 $R_\oplus$ away from us.

Structural naturalness, because of its subjective character, cannot be quantitatively defined. It is related to what the 1936 medicine Nobel laureate Henry Dale defines as {\it ``the subconscious reasoning which we call instinctive judgement"}~\cite{dale}. A more precise form of naturalness criterion has been developed in particle physics and it is playing a fundamental role in the formulation of theoretical predictions for new phenomena to be observed at the LHC. This criterion, called ``numerical naturalness" by Nelson~\cite{nelson}, will be the subject of this essay. 

\section{Drowning by Numbers} 
\begin{flushright}
{\it I am ill at these numbers.}\\
William Shakespeare~\cite{shake}
\end{flushright}
\label{secnum}

Our story starts with the observation that the ratio between the Fermi constant $G_F$ and the Newton constant $G_N$, which characterize respectively the strengths of the weak and gravitational forces, is a very large number\footnote{The figures in parenthesis give the one standard-deviation uncertainty in the last digits.}~\cite{pdg}
\beq
\frac{G_F\hbar^2}{G_Nc^2}=1.738~59(15) \times 10^{33}.
\label{gfgn}
\eeq
The powers of the Planck constant $\hbar$ and of the speed of light $c$ have been introduced in \eq{gfgn} to express the ratio as a pure number.

The human mind has always held in special fascination the pure numbers. Pythagoras went as far as believing that numbers are not just useful tools to describe the properties of nature but rather have special attributes that cause the various qualities of matter. Philolaus, a Pythagorean contemporary of Socrates and Democritus, expressed the idea that five is the cause of color, six of cold, seven of health, eight of love~\cite{hist}. These mystic properties of numbers are summarized in the motto of the Pythagorean school: {\it ``All is number"}. 

In a modern context, some numerical constants that appear in equations describing the fundamental laws of physics have often been the object of keen speculation. Sometimes these speculations are mere numerological exercises, but occasionally they are rewarded by a true understanding of deeper physical laws. When in 1885 Balmer first derived~\cite{balmer} a simple formula fitting the data for the frequencies $\nu$ of the hydrogen spectral lines
\beq
\nu = R \left( \frac{1}{n^2}-\frac{1}{m^2} \right) ~~~~{\rm with}~m>n~{\rm integers},
\eeq
he expressed bewilderment for 
{\it ``agreement which must surprise to the highest degree"}~\cite{balmer2}, but little did he suspect that Bohr's quantum interpretation~\cite{bohr} was lurking behind it. 

There are, however, less fortunate examples. From the very early times of electromagnetism and quantum mechanics, it was immediately recognized the special role of the fine-structure constant $\alpha$, a pure number constructed out of several fundamental quantities~\cite{pdg}
\beq
\alpha^{-1}=  \frac{4\pi \epsilon_0 \hbar c}{e^2}=137.035~999~11(46).
\eeq
Given its importance, there has been no lack of attempts to ``derive" $\alpha$ with simple numerical expressions. Early measurements were not even incompatible with the belief that $\alpha^{-1}$ must be an integer~\cite{eddinalp}. The hope was that finding the right formula for $\alpha$ would have opened the door towards a new theory underlying quantum electrodynamics, and curiously accurate expressions are, among many, $\alpha^{-1}=(8\pi^4/9)(2^45!/\pi^5)^{1/4}$~\cite{alp1}, $\alpha^{-1}=108\pi(8/1843)^{1/6}$~\cite{alp2}, $\alpha^{-1}=2^{-19/4}3^{10/3}5^{17/4}\pi^{-2}$~\cite{alp3}, $\alpha^{-1}=(137^2+\pi^2)^{1/2}$~\cite{alp4}. Even Heisenberg apparently took part in the game, with a less accurate try, $\alpha^{-1}=2^43^3/\pi$~\cite{alp5}. But, alas, these attempts are not particularly illuminating. Actually, a conceptual derivation of the fine-structure constant can be done in the context of grand unification, but the formula for $\alpha$ is certainly no easy guess for amateur numerologists\footnote{The formula is
$$
\alpha = \alpha_s \frac{\sin^2\theta_W \left( b_1-b_3 \right) +\frac 35 \cos^2\theta_W \left( b_3-b_2 \right)}{\left( b_1-b_2 \right)} + \mbox{higher-order~terms}.
\nonumber
$$
Here, the fine-structure constant $\alpha$, the strong coupling constant $\alpha_s$ and the weak mixing angle $\theta_W$ are evaluated at the same renormalization scale and $b_{1,2,3}$ are the gauge $\beta$-function coefficients. Higher-order terms cannot be neglected to achieve a prediction that matches the experimental accuracy.}.

The reason why speculating on the values of the fundamental constants may be meaningful is the reductionist belief in the existence of an underlying theory in which all dimensionless parameters are determined and computable. Einstein was firmly convinced that all forces must have an ultimate unified description and he even speculated on the uniqueness of this fundamental theory, whose parameters are fixed in the only possible consistent way, with no deformations allowed: {\it ``What really interests me is whether God had any choice in the creation of the world; that is, whether the necessity of logical simplicity leaves any freedom at all"}~\cite{eins}.  
This reductionist belief has enjoyed a spectacular success during the last century, bringing physics from the state of disconnected subjects (mechanics, optics, electromagnetism, thermodynamics, etc.) into the unified description of the Standard Model which, with a handful of free parameters, can accurately predict the properties of matter from distances down to about $10^{-16}$~cm to the conditions of the universe one second after the big bang. Nevertheless, it is this handful of free parameters which still escapes our understanding, preventing the fulfillment of Einstein's program. The determination of the ratio between Fermi and Newton constants in \eq{gfgn} is part of this puzzle.

The striking feature of the ratio in \eq{gfgn} is that its numerical value is huge. If the free parameters of the elementary-particle Standard Model are ultimately derived from a more fundamental theory,  they may carry information about deeper laws of physics.  What we observe as constants of order unity in the Standard Model could have a well-defined mathematical expression, in the more fundamental theory, containing numbers like $2$, $\pi$ or the like\footnote{My considerations here refer only to constants which are given by pure numbers; dimensionful constants define the units of measure.}.  On the other hand, if the constant is measured to be equal to a very large number, its ultimate expression cannot be a simple combination of 2's and $\pi$'s and we are inclined to think  that some important properties of the final theory can be learnt from its value. 

The lure of very large numbers is especially addicting. Eddington was stricken by the thought that the number of protons (equal to the number of electrons) in the universe, which he computed~\cite{eddington} to be equal to something like $10^{80}$, must be an exact {\it integer} number $N_E$. He was convinced that $N_E$ was not an accidental peculiarity of our universe, but rather a fundamental constant of nature. From this he deduced that the gravitational force between an electron and a proton ($G_Nm_em_p/r^2$) in a system of $N_E$ particles is given by the statistical fluctuation ($\sqrt{N_E}$) of the electric force between the two particles ($e^2/r^2$) and therefore~\cite{eddington2}
\beq
\frac{e^2}{G_Nm_em_p} =\sqrt{N_E}.
\eeq 
For $N_E=10^{80}$, this well agrees with the measured value ${e^2}/{G_Nm_em_p}=2.85\times 10^{40}$. To modern readers (and actually to many of his contemporaries as well) this argument has too much of a kabbalistic flavor. Nevertheless, it inspired Dirac to make his Large Number Hypothesis~\cite{lnh}. Any very large number occurring in nature should be simply related to a single very large number, which he chose to be the age of the universe. Indeed, he constructed three dimensionless numbers which all happen to be very close to $10^{40}$: the ratio of the size of the observable universe to the electron radius, the ratio of electromagnetic-to-gravitational force between protons and electrons, and the square root of the number of protons in the observable universe. To satisfy the Large Number Hypothesis, the ratio between any of these three numbers should remain roughly constant during the expansion of the universe. This can be achieved only if some fundamental constants vary with time, in order to maintain the proportionality of the three numbers. From this Dirac argued that the Newton constant $G_N$ should vary during the evolution of the universe, and he predicted its time dependence. This startling result and the fact that Dirac's paper was written during his honeymoon prompted Bohr's remark: {\it ``Look what happens to people when they get married!"}~\cite{gamow}. Indeed, Dirac's prediction was not very successful. His modification of gravity in the past would have changed the energy output of the Sun such that the oceans would have boiled in the pre-Cambrian era, while in fact life developed on Earth much earlier~\cite{teller}.

One lesson that we can learn from Dirac's hypothesis is that the existence of large numbers in nature may have nothing to do with the properties of the fundamental theory, but rather are the result of the cosmological history of our universe. Actually, as was first pointed out by Dicke~\cite{dicke}, the largeness of the three numbers examined by Dirac has a very simple explanation, which does not require any time-varying Newton constant. In order to reach the biochemical complexity that we observe on Earth, it is necessary for the universe to produce carbon, nitrogen, oxygen and other heavy elements which are synthesized in main-sequence stellar evolution and then dispersed throughout space by supernova explosions. An estimate of the time required by these processes, together with the information that the universe expands, shows that the three numbers considered by Dirac should indeed be at least as large as we observe them. Actually, they couldn't be much larger either, because otherwise hydrogen-burning stars, like our Sun, would have all burnt out. This means that we should have expressed surprise if Dirac's numbers had turned out to be of order one or much bigger than what they are, but their actual values lie indeed in the most reasonable range. A vast and old universe is an inevitable consequence of having observers like us. It is just a matter of the observer's point of view: although on Earth the Chinese are a million times more common than Mount Athos' inhabitants, if you happen to wonder around the Greek peninsula's monasteries, you will not be surprised to know that you have a much larger probability to encounter an orthodox monk rather than a Chinese person. In short, Dirac's problem appears as a red herring.

Can it be that also the $G_F/G_N$ ratio in \eq{gfgn} is large because of cosmological evolution or because of statistical probability, but carries no information whatsoever of the theory beyond the Standard Model?  I will come back to this question later, but for the moment it is more urgent to understand why the largeness of the number in \eq{gfgn} has anything to do with collider experiments at the LHC.

\section{A Quantum Complication}
\begin{flushright}
{\it Anyone who is not shocked by quantum\\ theory has not understood a single word.}\\
Niels Bohr~\cite{bohrquo}
\end{flushright}
\label{secqua}

The really problematic aspect about the $G_F/G_N$ ratio in \eq{gfgn} comes about when we consider the effects of quantum mechanics. In a quantum theory, the vacuum is a very busy place. Particle-antiparticle pairs are constantly produced out of nothing, violating the energy-conservation law by borrowing an amount of energy $E$ from the vacuum for a time $t$ such that $E~t<\hbar$, according to Heisenberg's uncertainty principle. These ``virtual" particles created from the vacuum have the same quantum numbers and properties as ordinary particles, with the exception that their energy-momentum relation is unusual ($E^2-p^2\ne m^2$). In the Standard Model, the size of $G_F$ is determined (up to coefficients which are unimportant for our discussion) by the mass of the Higgs boson $m_H$ , according to the relation $G_F \sim m_H^{-2}$. As the Higgs boson propagates in the quantum vacuum, it feels the presence of virtual particles and interacts with them. A characteristic property of the Higgs boson is to interact with any Standard Model particle with a strength proportional to the corresponding particle   
mass. Indeed, as Lenin once explained, {\it ``The Higgs mechanism is just a reincarnation of the Communist Party: it controls the masses"}~\cite{lenin}. When virtual particles appear in the vacuum,
they interact with the Higgs boson with an effective strength determined by the available energy $E$. Because of quantum corrections, the motion of the Higgs boson in the vacuum populated by virtual particles is affected by an amount proportional to $E$. As a result, the Higgs-boson squared mass $m_H^2$ receives an additional contribution
\beq
\delta m_H^2  = \kappa \Lambda^2,
\label{emax}
\eeq
where $\Lambda$ is the maximum energy $E$ accessible to virtual particles and $\kappa$ is a proportionality constant, which is typically\footnote{The contribution to $\kappa$ coming from virtual particles with the quantum numbers of the Standard Model degrees of freedom will be given in sect.~\ref{sec6}, see \eq{higgs}. It amounts to $\kappa =3\times 10^{-2}$.} in the range of $10^{-2}$.

A simple analogy can help us understand the result in \eq{emax}. Let us replace the quantum fluctuations of the vacuum with the more familiar thermal fluctuations of a thermodynamic system of a large number of particles at a temperature $T$. The particles (which I will call $P$) in this thermal bath play the role of the virtual particles in the quantum vacuum, and $T$ the role of the maximum available energy $\Lambda$. Let us now insert inside the box containing this hot $P$-particle gas a different particle initially at rest. I will call it $H$, as it plays the role of the Higgs in my analogy. At some initial time, $H$ has zero velocity and therefore its energy is equal to its mass, which I take it to be much smaller than the temperature ($E_H=m_H\ll T$). However, by statistical-mechanics arguments, we expect that the collisions of the particles $P$ will soon bring $H$ in thermal equilibrium, and therefore its energy will quickly become of order $T$. This is very similar to what happens in the quantum system, where the Higgs mass is pushed towards $\Lambda$, because of quantum-fluctuation effects.

The disturbing aspect of \eq{emax} is that it predicts that the Higgs mass $m_H$ ($\sim G_F^{-1/2}$) should be close to the maximum energy allowed by the theory. If the maximum energy is equal to the Planck mass $\mp$ ($=G_N^{-1/2}$), we find that the ratio $G_F/G_N$ is predicted to be rather close to unity, in strong contradiction with the measured value of $10^{33}$, see \eq{gfgn}.

One possible way out of the puzzle introduced by \eq{emax} is to assume that, once we include all quantum effects, the coefficient $\kappa$ in \eq{emax} is incredibly smaller than its typical value of $10^{-2}$. This requires a very precise cancellation of the different contributions to $m_H$ coming from different virtual particles at different energy scales. For instance, if we take $\Lambda =\mp$, the cancellation in $\kappa$ must be one part in $10^{32}$. This could occur just accidentally, as a result of the particular values chosen by nature for all the numerical constants entering in particle physics. But a purely fortuitous cancellation at the level of $10^{32}$, although not logically excluded, appears to us as disturbingly contrived. This is not what Einstein had in mind when he imagined a theory in which logical simplicity leaves no freedom at all. 

Just to get a feeling of the level of parameter tuning required, let me make a simple analogy. Balancing on a table a pencil on its tip is a subtle art that requires patience and a steady hand. It is a matter of fine tuning the position of the pencil such that its center of mass falls within the surface of its tip. If $R$ is the length of the pencil and $r$ the radius of the tip surface, the needed accuracy is of the order of $r^2/R^2$. Let us now compare this with the fine tuning in $\kappa$. The necessary accuracy to reproduce $G_F/G_N$ is equal to the accuracy needed to balance a pencil as long as the solar system on a tip a millimeter wide!     

This has led to a widespread belief among particle physicists that such an apparently fantastic coincidence must have some hidden reason. If we do not appeal to any special cancellation and fix $\kappa$ to its expected value of $10^{-2}$, then we can use \eq{emax} to extract the maximum energy up to which we can extrapolate our present knowledge of particle physics, and we find $\Lambda \approx$~TeV. Beyond the TeV a new theory should set in, modifying the Higgs mass sensitivity to quantum corrections. The LHC experiments, by studying particle collisions at energies above the TeV, will explore this new energy regime and will be able to tell us if the Standard Model is replaced by a new theory.

\section{The Naturalness Criterion as a Principle}
\begin{flushright}
{\it I have never lived on principles.}\\
Otto von Bismark 
\end{flushright}

\label{secnat}

We are now ready to formulate the naturalness criterion. Let us consider a theory valid up to a maximum energy $\Lambda$ and make all its parameters dimensionless by measuring them in units\footnote{Here I am following the usual convention of setting $\hbar =c=1$.} of $\Lambda$. The naturalness criterion states that one such parameter is allowed to be much smaller than unity only if setting it to zero increases the symmetry of the theory~\cite{thooft}. If this does not happen, the theory is unnatural.

There are two fundamental concepts that enter this formulation of the naturalness criterion: symmetry and effective theories. Both concepts have played a pivotal role in the reductionist approach that has successfully led to the understanding of fundamental forces through the Standard Model.

In modern physics, symmetries are viewed as fundamental requirements that dictate physical laws. If a parameter of the theory is equal to zero because of a symmetry, it will remain zero even after we have included all quantum corrections\footnote{Anomalous symmetries are exceptions to this rule, but they are not relevant to our discussion.}. This is why a small parameter is not necessarily problematic, if it is ``protected" by a symmetry according to the naturalness criterion stated above.

In the Standard Model there is no symmetry protecting the Higgs mass and this is the basic cause of the large quantum corrections in \eq{emax} that bring $m_H$ close to $\Lambda$. The absence of a symmetry protecting $m_H$ is linked to the spin-zero nature of the Higgs boson, as can be understood by a simple argument. Massless particles of spin $1/2$ or higher have two degrees of freedom. Massive particles of spin\footnote{Spin-$1/2$ Majorana particles are an exception. However, the symmetry argument applies also to this case, since the Majorana mass term violates the associated fermion number.}  $1/2$
 or higher have more than two degrees of freedom\footnote{This difference between massless and massive particles can be intuitively understood. A photon has two polarizations, the transverse modes along the direction of motion. But for a massive spin-1 particle, we can go to a reference frame where the particle is at rest. In that frame, we cannot distinguish between transverse and longitudinal modes, and therefore rotational invariance requires the existence of three polarization states. An analogous argument is valid for the spin-$1/2$ case. A massless spin-$1/2$ particle has a definite chirality. However, for a massive particle, with a boost along the direction of motion we can go to a frame where the chirality is opposite. Therefore relativistic invariance requires the massive particle to possess both chirality states. The argument cannot be repeated for a spin-0 particle, because there is no direction intrinsically defined by the particle itself.}. Therefore there is a conceptual distinction between the massless and massive cases. This distinction is due to the presence of an extra symmetry in the massless theory (gauge symmetry for spin~1, chiral symmetry for spin~1/2). The symmetry allows us to eliminate some degrees of freedom from the massless theory. This argument is valid for any particle with spin~1/2 or higher, but not for spin~0. There exist special symmetries able to protect spin-0 masses (non-linearly realized symmetries, supersymmetry) but they are not present in the Standard Model. This is why the Higgs boson is viewed as ``unnatural".
 
The second ingredient of the naturalness criterion is the use of effective field theories~\cite{effective}. Effective field theories are an extremely powerful concept. The idea is that, in a quantum field theory, it is possible to compute any physical process involving particles with momenta smaller than a maximum scale $\Lambda$ by replacing the original theory with a truncated version of it. This effective theory is expressed in terms of local operators that involve only light degrees of freedom. This means that the dynamics of low energies (large distances) can be fully described and computed by encoding the information of high energies (small distances) into a finite number of parameters. Effective field theories are a powerful realization of the reductionist approach. As we increase the distance scale, we increase the complexity of the system and new phenomena emerge. These phenomena are best described by an effective theory, for which knowledge of the full details of the underlying theory is unnecessary, but can be summarized in a finite number of parameters. These parameters can be experimentally measured or theoretically derived (and possibly both). The way thermodynamics can be derived from statistical mechanics is a good example of this reductive process.

The naturalness criterion, as stated above, excludes the possibility that the parameters that encode the information of physics at very short distances are correlated with dynamics of the effective theory occurring at large distances. Such a correlation would signal a breakdown of the philosophy underlying the effective-theory approach\footnote{This would not mean that the effective-theory approach is useless. It would only mean that certain properties of the theory cannot be captured by low-energy arguments alone. The conjecture of gravity as the weakest force~\cite{graweak}, if true, is one example of a theoretical property that cannot be derived using an effective-theory approach.}. If the naturalness criterion is a good guiding principle, we expect to discover new particles at the LHC, associated to the taming of the Higgs-mass quantum corrections. Some theoretical proposals that describe these new particles are discussed in other chapters of this book~\cite{nima,savas}.
If experiments at the LHC find no new phenomena linked to the TeV scale, the naturalness criterion would fail and the explanation of the hierarchy $G_F/G_N$ would be beyond the reach of effective field theories.

\section{An Account of Events}
\begin{flushright}
{\it History is a set of lies agreed upon.}\\
Napol\'eon Bonaparte 
\end{flushright}

The concept of naturalness and its implications for electroweak physics did not spring from a single paper but, rather, they developed through a ``collective motion" of the community which increasingly emphasized their relevance to the existence of physics beyond the Standard Model. I will give here a short account of how the naturalness criterion for the Higgs boson mass was developed by theoretical particle physicists. 

Starting in 1976, the work by Gildener and Weinberg~\cite{gildener} revealed a conceptual difficulty with the recently discovered grand unified theories, the so-called hierarchy problem. One-loop quantum corrections were found to give contributions to the Higgs mass proportional to the mass of the superheavy states, of the order of $M_{GUT}=10^{14-16}$~GeV. Keeping a hierarchical separation of scales between $M_W$ and $M_{GUT}$ required fine tuning the parameters of the theory of more than $10^{-24}$. This is nothing less than a specific realization of the Higgs naturalness problem, in the presence of a theory with two widely separated scales. Even today some people find it easier to understand and to accept the naturalness problem in this context, since one makes no reference to cut-off (and regularization procedure) dependent quantities of the effective theory\footnote{Shaposhnikov~\cite{shap} concedes that there is a Higgs naturalness problem in presence of $M_{GUT}$, but he argues that in the absence of any new mass scale between the weak and the Planck scale the problem may not exist since, according to him, the Planck mass could be conceptually different from the field-theoretical ultraviolet cutoff of the effective low-energy theory.}.

In 1978, Susskind~\cite{susskind} introduced the naturalness problem of the Higgs as a primary motivation for his proposal of technicolor, giving however full credit to Wilson for pointing out the conceptual difficulty linked to the existence of fundamental scalar particles. Indeed, in an article written at the end of 1970, Wilson had clearly expressed the problem, from an effective-theory point of view: {\it``It is interesting to note that there are no weakly coupled scalar particles in nature; scalar particles are the only kind of free particles whose mass term does not break either an internal or a gauge symmetry. This discussion can be summarized by saying that mass or symmetry-breaking terms must be ``protected" from large corrections at large momenta due to various interactions (electromagnetic, weak, or strong). A symmetry-breaking term $h_\lambda$ is protected if, in the renormalization-group equation for $h_\lambda$, the right-hand side is proportional to $h_\lambda$ or other small coupling constants even when high-order strong, electromagnetic, or weak corrections are taken into account [\dots ]. This requirement means that weak interactions cannot be mediated by scalar particles"}~\cite{wilson1}. He could not have been more explicit. Nevertheless, in 2004 Wilson completely retracted, while recalling the results he obtained in the early 1970's: {\it ``The final blunder was a claim that scalar elementary particles were unlikely to occur in elementary particle physics at currently measurable energies [\dots ]. This claim makes no sense"}~\cite{wilson2}.

The naturalness criterion, in the way I stated it in sect.~\ref{secnat}, was formulated by 't~Hooft in lectures held in 1979~\cite{thooft}. Actually a precursor of this criterion was Gell-Mann's totalitarian principle which states: {\it ``Everything which is not forbidden is compulsory"}\footnote{Although the totalitarian principle is indisputably attributed to Gell-Mann, I could not trace the original source. The earliest reference to it that I found is ref.~\cite{sudarshan}. In the first version of this essay I stated that the totalitarian principle's expression is borrowed from ``The Once and Future King" by T.H.~White, published in 1958. I thank Stanley Deser who pointed out to me that the expression is actually coming from ``Nineteen Eighty-Four" by G.~Orwell, published in 1949.}. It refers to the property, largely confirmed by experimental evidence, that every interaction term not explicitly forbidden by conservation laws must be present. Quantum corrections in an effective theory appear to enforce the totalitarian principle by giving large contributions to parameters that are not forbidden by a symmetry.

Although by 1979 the Higgs-naturalness problem had been clearly spelled out, supersymmetry as a possible solution is only mentioned in some lectures held by Maiani in that year: {\it ``In a supersymmetric theory, one could hope to obtain that the bare curvature of $V_{eff}$ vanishes and it is not renormalized by radiative corrections [\dots ] No concrete model of this type have been constructed yet"}~\cite{maiani}. Supersymmetric models were being developed for years, most notably by Fayet~\cite{fayet}, but with no connection to the naturalness problem. Although the non-renormalization theorems had already been discovered, supersymmetry was seen more as a way to unify gravity and gauge forces~\cite{zumin}, rather than a way to address the hierarchy problem. Probably many physicists did not attach great importance to the naturalness problem of the Higgs mass, simply because the Higgs model did not appear to be very compelling, as was expressed by Iliopoulos in the 1979 Einstein Symposium: {\it ``Several people believe, and I share this view, that the Higgs scheme is a convenient parametrization of our ignorance concerning the dynamics of spontaneous symmetry breaking and elementary scalar particles do not exist"}~\cite{iliopoulos}.

Things changed by 1981. At the end of 1980 Veltman had published an influential paper emphasizing the problem~\cite{veltman}. In 1981 Witten clearly pointed out how supersymmetry can solve the naturalness problem and explained the crucial role of dynamical supersymmetry breaking~\cite{witten}. About a month later Dimopoulos and Georgi~\cite{dimgeo}, using the results of Girardello and Grisaru on soft supersymmetry breaking~\cite{girardello}, developed a simple and realistic grand unified supersymmetric model. The age of supersymmetric model building had started and an explosion of activity followed. Since then, the Higgs naturalness problem has become one of the most studied puzzles in particle physics and one of the driving motivations to explore physics beyond the Standard Model. 

\section{The Paths Chosen by Nature}
\label{sec6}
\begin{flushright}
{\it Can we actually know the universe?\\ My God, it's hard enough finding\\ your way around in Chinatown.}\\
Woody Allen~\cite{allen}
\end{flushright}

How does nature deal with the hierarchy between $G_F$ and $G_N$? Does nature respect the naturalness criterion? Experiments at the LHC will be able to shed some light on these questions. In the meantime, we can only use our imagination. Something useful can be learned by studying how nature deals with other problems, which have similar characteristics, but for which we already know the answer. 

An interesting analogy was first suggested, to the best of my knowledge, by Murayama~\cite{mura}.
Consider the electron as a sphere of radius $r$. The electromagnetic energy associated with this configuration is $\alpha/r$. This energy must be smaller than the total energy of the electron, equal to $m_ec^2$, where $m_e$ is the electron mass. Therefore, we obtain
\beq
r > \frac{\alpha}{m_e}=3\times 10^{-15}~{\rm m}.
\eeq
In words, the electron radius has to be larger than an atomic nucleus! Things get even worse when we include the magnetic energy of a spinning sphere $\mu^2/r^3$ (where $\mu=e\hbar /(2m_e c)$ is the electron magnetic moment), as done by Rasetti and Fermi~\cite{rasetti}, immediately after the discovery of the electron spin. In this case, one finds $r>\alpha^{1/3}/m_e$.

The puzzle is the following. Either the different contributions to the total electron energy mysteriously cancel with a high precision, or some new physics sets in before the energy scale $r^{-1}\sim m_e/\alpha$, modifying the electromagnetic contribution to the electron mass at short distances and preserving naturalness. 
In this example, nature has chosen the second option. Indeed Dirac showed that a new particle with mass $m_e$, the positron, has to be included in a consistent relativistic quantum theory. As explicitly calculated by Weisskopf~\cite{weisskopf}, the electromagnetic contribution to the electron mass at small distances grows neither like $1/r$ nor like $1/r^3$, but rather like $\alpha~ m_e\ln (m_e r)$. This contribution is less than the electron mass even for distances $r$ as small as the Planck length. In this case, nature has preferred to obey the naturalness criterion.

There are several other examples one can consider where physical quantities computed in the effective theory require either cancellations of contributions sensitive to the small-distance regime, or the appearance of new physics that restore naturalness. In many cases, nature has chosen to preserve naturalness and new particles at the appropriate energy scale modify the theory. For instance, the electromagnetic contribution to the charged to neutral pion mass difference is
\beq
M^2_{\pi^+}-M^2_{\pi^0}= \frac{3\alpha}{4\pi}\Lambda^2,
\label{pion}
\eeq
where $\Lambda$ is the ultraviolet momentum cutoff, {\it i.e.} the maximum energy of the effective theory of pions. The request that \eq{pion} not exceed the measured quantity $M^2_{\pi^+}-M^2_{\pi^0}=(35.5~{\rm MeV})^2$, implies that $\Lambda$ must be smaller than 850~MeV. Indeed, before that mass scale, the $\rho$ meson exists ($M_\rho = 770$~MeV) and the composite structure of the pion softens the electromagnetic contribution. 

Another example is the mixing between the $K^0$ and ${\bar K}^0$ mesons. The mass difference between the $K^0_L$ and $K^0_S$ states, as computed in an effective theory valid at energies of the order of the kaon mass, is given by
\beq
\frac{M_{K^0_L}-M_{K^0_S}}{M_{K^0_L}} =\frac{G_F^2f_K^2}{6\pi^2}~\sin^2\theta_c~\Lambda^2,
\label{kaon}
\eeq
where $f_K=114$~MeV is the kaon decay constant and $\sin\theta_c=0.22$ is the Cabibbo angle.
If we require that the result in \eq{kaon} be smaller than the measured value $(M_{K^0_L}-M_{K^0_S})/M_{K^0_L}=7\times 10^{-15}$, we find $\Lambda < 2$~GeV. Indeed, before reaching this energy scale a new particle (the charm quark with mass $m_c\approx1.2$~GeV) modifies the short-distance behavior of the theory, implementing the so-called GIM mechanism~\cite{gim}. Incidentally,
while the other two examples are {\it a posteriori} deductions, the case of $K^0$--${\bar K}^0$ mixing is historically accurate: this is the actual argument used by Gaillard and Lee~\cite{lee} to compute the mass of the charm quark before its discovery.

We can formulate the problem of the Higgs mass $m_H$ in the same fashion. Using the Standard Model as an effective theory, we can compute the contributions to $m_H$ due to Higgs interactions. The leading effect is
\beq
\delta m_H^2 =\frac{3G_F}{4\sqrt{2}\pi^2}\left( 4m_t^2-2m_W^2-m_Z^2-m_H^2\right) \Lambda^2,
\label{higgs}
\eeq
where $m_t$, $m_W$, $m_Z$ are the masses of the top quark, $W$ and $Z$ gauge bosons, and $\Lambda$ is the maximum momentum\footnote{Naively one may think that the Higgs naturalness problem disappears for the special value of $m_H$ that cancels the right-hand side of \eq{higgs} (which happens to be about 200--300~GeV, depending on the value of the renormalization scale). Unfortunately this is not sufficient because \eq{higgs} gives only the infrared contribution to $m_H$. Modes with masses of order $\Lambda$ (outside the domain of the effective theory) give new contributions of the same size. For example, in a softly-broken supersymmetric theory, quadratic divergences are absent, but this is not sufficient to solve the hierarchy problem. It is also necessary that the masses of the new particles lie below the TeV scale.}.  The request that the contribution in \eq{higgs} be not larger than 182~GeV (the 95\% CL limit from Standard Model fits of present experimental data~\cite{lepew}), implies $\Lambda <1.0$~TeV. Only the LHC will tell us if the naturalness criterion is successful in this case as well, and whether new particles exist with masses below the TeV.

Unfortunately not all examples are successful and there is one important case in which nature does not seem to respect the naturalness criterion. Astronomical observations place bounds on the energy density of the vacuum in our universe which constrain the scale of the cosmological constant to be less than  $3\times 10^{-3}$~eV. Since quantum corrections to the cosmological constant grow with the maximum energy $\Lambda$, the naturalness criterion implies that our theoretical description of particle physics should start failing at an energy scale as low as $3\times 10^{-3}$~eV. We have good evidence that this is not the case. Nature could have chosen supersymmetry to deal with this problem in a {\it natural} way because the cosmological constant vanishes in supersymmetric theories. However, we already know that nature has decided not to take this opportunity, since supersymmetry is not  an exact symmetry down to energies of  $3\times 10^{-3}$~eV.

The issue is more involved, because the cosmological constant becomes a physical observable only when we include gravity, which can be usually ignored when dealing with particle physics processes. If a solution to the cosmological constant exists, it may involve some complicated interplay between infrared and ultraviolet effects (maybe in the context of quantum gravity) or it may just be linked to the cosmological history. At any rate, none of these solutions will be obtained by an effective field theory approach. But then, are we sure that this is not the case also for the Higgs mass? The verdict will be handed down by the LHC. 

\section{Measuring Naturalness}
\begin{flushright}
{\it I used to measure the heavens,\\ now I measure the shadows of earth.}\\
Johannes Kepler~\cite{kepler}
\end{flushright}

As new particle physics theories were invented to cope with the naturalness problem of the Higgs mass, and as collider experiments started to set bounds on the existence of the new particles, there was a need to give a quantitative criterion for the degree of naturalness (or unnaturalness) of the new theories. A commonly adopted criterion~\cite{bargiu} was to consider the expression of the $Z$ boson mass (which is equivalent, up to constants of order unity, to $m_H$ or to $G_F^{-1/2}$) as a function of the parameters $a_i$ of the underlying theory. Indeed, such an expression should always exist, since in the new theory the weak scale must be a ``calculable" quantity (although calculable only in terms of unknown parameters).  The measure of naturalness (or, more precisely, of the amount of fine-tuning) is given by the logarithmic variation of the function $M_Z (a_i)$ with respect to $a_i$,
\beq
\Delta \equiv {\rm max}\left| \frac{a_i ~\partial M_Z^2(a_i)}{M_Z^2~\partial a_i}\right| .
\label{crit}
\eeq
A theory with $\Delta =10$ suffers from a parameter tuning of no more than 10\%, one with $\Delta =100$ of {1\%}, and so on. 

For example, in the case of supersymmetry, the requirement of less than 10\% tuning led to the prediction that supersymmetry had to be discovered at LEP2. This prediction turned out to be wrong. Indeed, today supersymmetric models pass the experimental tests only if their free parameters are tuned at the level of few percent. Actually this is essentially true for all known extensions of the Standard Model that address the Higgs mass problem. Of course, one can argue that the Sun and the Moon have radius and distance from the Earth  ``tuned" to appear equal in the sky (with a precision of about 5\%), for no better reason than producing rare and spectacular eclipses (and permitting us to test general relativity). Even more dramatic numerical coincidences happen in nature. Still, I would  hope that the new theory of electroweak interactions, whatever that is, ``naturally" solves the naturalness problem. 

It may well be that, in some cases, \eq{crit} overestimates the amount of tuning. Indeed, \eq{crit} measures the sensitivity of the prediction of $M_Z$ as we vary parameters in ``theory space". However, we have no idea how this ``theory space" looks like, and the procedure of independently varying all parameters may be too simple-minded\footnote{For instance, some authors have argued that, supersymmetric models become less fine-tuned if one imposes special restrictions on the theoretical parameters at the GUT scale (like ${\tilde m}_t={\tilde m}_H$ and large $\tan\beta$~\cite{feng} or ${\tilde m}_t^2\approx -4M_{\tilde g}^2$~\cite{kim}). In the absence of solid theoretical motivations for these restrictions, it is difficult to assess the real benefits of such approaches.}. In conclusion, although a quantitative measure of naturalness can be of useful guidance to build new theories, it is very easy to slip into purely academic and sterile considerations. As we are drawing closer to the beginning of LHC operations, the real issue is whether the new theory predicts observable phenomena in the TeV domain or not.
 
\section{Anthropic Reasoning}
\begin{flushright}
{\it A physicist talking about the anthropic principle runs\\ the same risk as a cleric talking about pornography:\\ no matter how much you say you are against it,\\ some people will think you are a little too interested.}\\
Steven Weinberg 
\end{flushright}

Is the naturalness of the Higgs mass a good scientific question that will make us understand fundamental properties of nature? There are some questions that at first sight appear pregnant with deep meanings, but then end up to be red herrings. Probably Dirac's question (``Why are these numbers so large?") was one of them because, as we have seen in sect.~\ref{secnum}, his explanation in terms of a time-varying $G_N$ was less successful than Dicke's simple observation based on the essential role of contingency in the observation. An alien landing on Mount Athos is warned: do not make wrong conclusions on the mystical inclinations of earthlings, before carefully considering the circumstances of your observation.

In 1595 Kepler asked the apparently good scientific question ``Why are there six planets?", and in {\it Mysterium Cosmographicum} proposed an attractive symmetry-based answer. Planetary orbits lie on successive spheres that circumscribe and inscribe the five Platonic solids\footnote{It is interesting to note how the number of space dimensions plays an essential role in this hypothesis. In three dimensions there exist only five regular solids but, in two dimensions, there is an infinite number of regular polygons, and therefore an infinite number of planets.}. Based on this hypothesis he could predict the ratio of the planetary distances, which matched observations well within the accuracy known at the time. Of course today we known that the number of planets and their distances from the Sun do not carry any significant information on the fundamental laws of physics; hence, another red herring.

Even from these ``wrong" questions there is a lesson to be learned. Special incidents may not be an indication of some deep property of the fundamental theory, but just the consequence of the special condition of the observer~\cite{anthrop}. However, for this to happen, there must exist a large ensemble of possible incidents, from which the special observer picks a special case. In practice this means that, if we do not want to attach a special significance to our observation, we learn something about the ensemble. From large numbers, we deduce that the universe must expand; from meeting a thousand Orthodox monks, we conclude that the Earth is highly populated; from the special location of the Earth in the solar system, we deduce that the universe must contain a large number of stars.

In the same way, the measured value of $G_F/G_N$, which seems special to us, could actually be a very plausible observation in a universe that has developed complex structures,  if there exists a multitude of universes with different values of $G_F/G_N$. In the vast majority of the universes $G_F/G_N$ is of order unity, but those universes do not have the right properties to develop observers. Indeed, the measured value of $G_N$ appears very favorably chosen to sustain non-trivial chemistry~\cite{donog} (the same can be said about the cosmological constant, since the existence of galaxies is very sensitive to its value~\cite{weincc}). This picture of a multitude of parallel universes, usually referred to as the ``{\it multi}verse" (as opposed to a single {\it uni}verse), can be realized in the context of string theory and eternal inflation~\cite{multi}. If true, it would represent the next step in Copernican revolution: not only is the Earth not special, but even the universe in which we live is just one out of many.

Does this scenario imply that the Higgs naturalness problem was a red herring and that the LHC is doomed to find the Higgs particle and nothing else? Quite possible. However, sometimes there are remarkable properties that unexpectedly emerge. Sometimes they are simple coincidences, but sometimes they hide significance of great importance. A most singular episode is related by Barrow~\cite{barrow}. Unattested tradition narrates that William Shakespeare may have contributed to the English renderings of the Psalms in the King James Version of the Bible. An Eton schoolboy noticed that in Psalm 46, written in the year in which Shakespeare (who was born in 1546) was 46 years old, the word ``SHAKE" is the 46th from the beginning, and ``SPEAR" is the 46th from the end. Coincidence or a hidden signature of the poet?

Supersymmetry at the weak scale was introduced to tame the quantum corrections to the Higgs mass. However, it has been noticed that the supersymmetric particles have exactly the right quantum numbers to unify the gauge couplings at a very large energy scale with surprising precision. Moreover, the massive, neutral, stable Majorana particle that automatically emerges from many supersymmetric theories is exactly what is needed to account for the dark matter observed in our universe. Coincidences or hidden signatures of supersymmetry?

These observations have led to the proposal of Split Supersymmetry~\cite{split}, in which gauge-coupling unification and dark matter are taken as basic elements, while the solution of the Higgs naturalness problem is abandoned. This theory has several interesting features and quite distinctive signatures at collider experiments. If confirmed by the LHC, it would provide tangible experimental evidence against the naturalness criterion.  

\section{Naturalness versus Criticality}
\begin{flushright}
{\it Results without causes are much more impressive.}\\
Sherlock Holmes~\cite{holmes1} 
\end{flushright}

There is a different way of looking at the hierarchy problem $G_F/G_N$. In the Standard Model the weak scale is determined by the vacuum expectation value of the Higgs field, which triggers electroweak symmetry breaking. The order parameter of the phase transition can be expressed in terms of the coefficient $\mu^2$ that enters the Higgs potential. If $\mu^2$ is positive the symmetry remains unbroken, if $\mu^2$ is negative the symmetry is broken, and $\mu^2=0$ defines the critical point. This is completely analogous to the Ginzburg-Landau description of ferromagnetism. For temperatures $T$ larger than the critical Curie temperature $T_C$, the dipoles are randomly oriented, the total magnetization vanishes, and the system is rotationally symmetric. When $T-T_C$ becomes negative, the dipoles are aligned creating a spontaneous magnetization, and the system breaks rotational symmetry.

Because of quantum corrections, we expect $|\mu^2|$ to be close to the maximum energy $\Lambda^2$ and, depending on its sign, to break or preserve electroweak symmetry.  The hierarchy problem can then be rephrased in the following way~\cite{giurat}: if the critical value separating the two phases is not special from the point of view of the fundamental theory, why are the parameters in the real world chosen such that we live so near the critical condition?

There are systems in nature which have the tendency to evolve into critical states, even if there is no outside agent that forces them in that direction. This process is called self-organized criticality~\cite{soc}. The prototype example is a sand pile where grains of sand are slowly added. As the pile grows, it reaches a condition where catastrophic sand slides occur after the addition of just a single grain. Avalanches of all sizes obey a power-law distribution and therefore the dynamics of the system can no longer be understood in terms of single grains. There are correlations among distances vastly larger than the size of the grain of sand. The system has arranged itself to be near critical and remains close to the critical condition (as long as we continue to slowly add more sand). There are many, apparently unrelated, phenomena that seem to follow this pattern: from the distribution of earthquake intensity to extinctions of biological species; from river bifurcations to traffic jams. 

Is it possible that a pattern of self-organized criticality with respect to electroweak symmetry brings the Standard Model towards the condition of a large hierarchy $G_F/G_N$? If anything like this operates in nature, then it will not be captured by an effective-theory approach and it will not respect the naturalness criterion.  The microphysics description will fail to properly account for some large-scale correlations, in the same way as individual grains are not useful to describe the avalanches in the sand pile occurring at all scales (between the size of a single grain and the size of the whole pile). To realize such an idea, an ensemble of theories seems to be a necessary ingredient, and therefore we still have to rely on the multiverse. However, the process of selection of our universe will be, in this case, determined by dynamics rather than by anthropic considerations.

\section{Conclusions}
\begin{flushright}
{\it ``Data! Data! Data!" he cried impatiently.\\ ``I can't make bricks without clay".}\\
Sherlock Holmes~\cite{holmes2} 
\end{flushright}

The primary goal of the LHC is to discover the mechanism of electroweak symmetry breaking. Indeed, the Standard Model, including only the particles known today, becomes inconsistent at an energy scale of about 1~TeV. The LHC, producing particle collisions with energies above this scale, is bound to probe the mechanism of electroweak breaking, whether it is given by the Higgs or by some alternative dynamics.

There is a second, more subtle, issue related to the existence of a fundamental Higgs boson, which will also be investigated by the LHC. The basic problem is the absence, within the Standard Model, of symmetries protecting the Higgs mass term, and therefore the expectation that the maximum energy up to which the theory can be {\it naturally} extrapolated is, again, the TeV. A new physics regime should set in at that energy scale, and the hypothetical Higgs boson must be accompanied by new particles associated with the cancellation of the quantum corrections to $m_H$. This is not a problem of internal consistency of the theory, but an acute problem of naturalness. As such, it does not necessarily guarantee that a new physics threshold really exists in nature. But, if new particles at the TeV scale are indeed discovered, it will be a triumph for our understanding of physics in terms of symmetries and effective field theories. 

This is, in conclusion, the naturalness problem that theoretical particle physics is facing today. If you found the subject too speculative, be reassured: time has come for the question to be settled by experimental data.

\end{document}